# On pitfalls (and advantages) of sophisticated Large Language Models

Anna Strasser [1,2]

[1]Faculty of Philosophy, Ludwig-Maximilians-Universität München

[2]DenkWerkstatt Berlin

**Abstract**

Natural language processing based on large language models (LLMs) is a booming field of AI research. After neural networks have proven to outperform humans in games and practical domains based on pattern recognition, we might stand now at a road junction where artificial entities might eventually enter the realm of human communication. However, this comes with serious risks. Due to the inherent limitations regarding the reliability of neural networks, overreliance on LLMs can have disruptive consequences. Since it will be increasingly difficult to distinguish between human-written and machine-generated text, one is confronted with new ethical challenges. This begins with the no longer undoubtedly verifiable human authorship and continues with various types of fraud, such as a new form of plagiarism. This also concerns the violation of privacy rights, the possibility of circulating counterfeits of humans, and, last but not least, it makes a massive spread of misinformation possible.

## 1. Introduction

Natural language processing based on large language models (LLMs), such as BERT, Eleuther, ChatGPT, GPT-3, LaMDA, and PALM, is a booming field of AI research. In general, artificial systems based on neural networks and deep learning have achieved impressive success in many domains based on pattern recognition, like speech recognition, lipreading, and game-playing. Examples in this area include LipNet (Assael et al. 2016), a program for lung cancer screening (Ardila et al. 2019), AlphaFold – predicting protein structure (Jumper et al. 2021), and AlphaTensor – discovering novel matrix multiplication algorithms (Fawzi et al. 2022), DeepBlue (Campbell 2002), AlphaGo and other game-playing programs (Brown & Sandholm 2019; Silver et al. 2016, 2018). With respect to language, we have observed impressive progress in automatic translation (DeepL) and computer code generation (GitHub Copilot) over the past few years, all relying on LLMs and deep learning. Remarkably, LLMs are able to produce grammatically correct linguistic outputs with fluency, often similar to that of a human. Many of their outputs can hardly be distinguished from linguistic outputs originally created by humans, even though some demonstrate a lack of common sense and embarrassingly expose the models.[1] After neural networks have proven to outperform humans in games and practical domains relying on pattern recognition, we potentially now stand at a road junction where artificial entities could eventually enter the realm of human communication.

However, this comes with serious risks that we should consider with caution. The widespread use of LLMs, along with expected advances in their development, will make it increasingly difficult to distinguish between human-written and machine-generated text. This fact alone is already going to create all sorts of new challenges. For instance, it will be difficult to prove human authorship beyond doubt, and it will be equally difficult to unmask with certainty machine-generated text that is fraudulently passed off as self-written text. Furthermore, text generated by LLMs has the potential to be abused for a new kind of plagiarism, fraud, and the spread of misinformation. Due to the inherent limitations regarding the reliability of neural networks, overreliance on LLMs can have disruptive consequences. Precisely because they are so good at mimicking human linguistic performance, there is a risk that they will be used to mass-produce misinformation (Marcus 2022, 2023). With the help of LLMs, it will be even easier than before to create an infinite amount of text for troll farms and fake websites, which in turn will lead to a decline in the level of trustworthiness on the Internet. Even though the outputs of LLMs often sound very convincing, they are not reliably truthful. Critical voices say that, for example, ChatGPT is a 'bullshit generator' (McQuillan 2023). One could go so far as to conjecture that with the further development of such language models, a new class of weapons is emerging that can have devastating effects in the war for truth (cp. Guardian editorial 2023).

---

[1] For examples of typical mistakes, see ChatGPT/LLM error tracker (Davies et al. 2023).

In this chapter, I focus on the consequences current and further advanced LLMs might have. I start with an overview of how difficult it is already now to distinguish between machine-generated and human-made text (*2. Difficult to distinguish*). Thereafter, I discuss various ethical consequences arising from this indistinguishability. Specifically, I address challenges related to difficult-to-prove human authorship, examine new forms of plagiarism, and critically evaluate copyright and privacy issues related to the model construction. I also consider the possibilities of counterfeiting people and spreading misinformation and toxic language (*3. Ethical consequences*). In the next section, I discuss possible ways by which society might deal with the inherent risks of LLMs. This concerns not only a possible adaptation of our legal basis but also the technical possibilities available for implementing the legislation (*4. How to handle the epistemological crisis*). After addressing at length the potential risks associated with the increased use of LLMs, I turn to the question of the extent to which there might also be helpful applications arising from LLMs that could be used in everyday life (*5. LLMs as thinking tools*).

## 1.1. Background

Since research related to LLMs is a fairly young and rapidly evolving field of research, it is difficult to provide an overview of the state of the art that is not already outdated at the time of publication, as there is a constantly growing body of new publications. Research papers, as well as opinion papers from computer science and philosophy and various publications in the media, served as the basis for this chapter. These are, for example, technical papers introducing specific LLMs such as BERT (Devlin et al. 2018; Rogers et al. 2020), Eleuther (https://www.eleuther.ai), GPT-3 (Brown et al. 2020), LaMDA (Thoppilan et al. 2022), and PALM (Chowdhery et al. 2022) as well as meta-reviews on benchmarks used to assess the performance of language models (Michael et al. 2022, Srivastava et al. 2022).

Since the initial release of GPT-3 on June 11, 2020, LLMs have attracted the interest of other disciplines, such as linguistics, cognitive science, philosophy, and others, and have also received considerable public attention (Mahowald et al. 2023; Marcus & Davis 2020; Heaven 2020; Simonite 2020; Weinberg (ed.) 2020). For example, Mahowald and colleagues (2023) investigate the capabilities of LLMs, distinguishing between formal competence (the knowledge of linguistic rules) and functional competence, which refers to understanding and using language in the world. They conclude that LLMs are close to mastering formal competence but fail at functional competence tasks. Public perception is initially blown away by the impressive achievements of LLMs, and also from scientific communities, the degree of impressiveness is immense as it is shown, e.g., by the title of an opinion piece by Will Douglas Heaven (2020): "OpenAI's new language generator GPT-3 is shockingly good – and completely mindless."

In June 2022, Google's LaMDA model made international headlines when Google engineer Blake Lemoine said he became convinced that LaMDA was sentient, prompting a flood of papers (Bryson 2022; Frankish 2022; Hofstadter 2022; Klein 2022; Roberts 2022; Shanahan 2023; Tiku 2022). Although the majority is not inclined to ascribe sentience to LLMs, there is a trend to use philosophically loaded terms, such as "knowing," "believing," "comprehending," and "thinking" when describing these systems. The extent to which this can be a justified use of these terms is open to debate.

Similarly, ChatGPT, launched in November 2022, evoked a long-lasting echo in the media and the academic communities (Chiang 2023; Krakauer & Mitchell 2022; Lock 2022; Roose 2022; Thorp 2023; Wolfram 2023). For example, Derek Thompson (2022) mentioned ChatGPT in *The Atlantic* magazine's "Breakthroughs of the Year" as part of "the generative-AI eruption" that "may change our mind about how we work, how we think, and what human creativity really is." In the beginning, the enthusiastic voices received a lot of attention, but now the critical voices seem also to gain more consideration (Guardian editorial 2023; Hofstadter 2022; Marcus 2020, 2023; Marcus & Davis 2020, 2023; McQuillan 2023).

## 2. Hard to distinguish

Just ten years ago, people didn't give much thought to how to distinguish machine-generated text from human-generated text. The differences were so obvious back then, and it didn't seem like that would change quickly. Back then, the differences were so obvious, and it didn't look like that that this would change anytime soon. However, with the advent of more and more upscaled LLMs, this is becoming a serious problem. Today, in social media, e-customer service, and advertising, we are increasingly exposed to machine-generated content that can easily be mistaken for human-generated content. Neither humans nor sophisticated detection software can distinguish with certainty between human-generated and machine-generated text. In empirical research, this indistinguishability, along with the tendency of humans to anthropomorphize, is sometimes even exploited when experimental protocols with artificial agents are used to test hypotheses about human social cognitive mechanisms (Strasser 2022; Wykowska et al. 2016). And it is already a concern for teachers that they will not be able to distinguish their students' self-written essays from machine-generated ones (Herman 2022; Hutson 2022; Huang 2023; Marche 2022; Peritz 2022 Sparrow 2022).

### 2.1. Human discrimination abilities

The more advanced LLMs are, the more difficult it becomes for humans to distinguish between machine-generated and human-made text. Besides various rather informal assessments (Rajnerowicz 2022; Sinapayen 2023; Vota 2020), there are three studies using rigorous psychological methods to test human's ability to distinguish between machine-generated and human-made text (Clark et al. 2021; Brown et al. 2020; Schwitzgebel et al. 2023).

By demonstrating a difference between the two basic models of Open-AI (GPT-2 and GPT-3), Clark and colleagues (2021) were able to show that the more advanced the LLMs, the more difficult the distinction becomes. They collected short human-generated texts in three domains: stories, news articles, and recipes, and used the two base models to generate texts within the same domains. The participants (6 groups covering the three domains for each model) were then presented with five selected texts and asked to judge whether these texts were likely to have been generated by humans or by machines. Results for the older model, GPT-2, showed that participants were able to accurately distinguish between GPT-2-generated and human-generated texts 58% of the time, significantly above the chance rate of 50%. In contrast, accuracy in discriminating between the newer model (GPT-3) and human-generated text was only 50%, not significantly different from chance. Even additional training in follow-up experiments failed to increase accuracy to above 57% in any domain. These results show that scaling up the models makes it more difficult to distinguish between human-made and machine-generated text, and it is expected that the results of such an experiment will point even more clearly in this direction when GPT-4 is on the market.

Brown et al. (2020) focused on the domain of news articles and found similar results, indicating a moderately good discrimination rate for smaller (older) language models and near-chance performance with the largest version of GPT-3.

In the study I conducted with Eric and David Schwitzgebel (Schwitzgebel et al. 2023), we fine-tuned the Davinci model of GPT-3 on the corpus of the well-known philosopher Daniel Dennett (Strasser et al. 2023) and tested three groups of participants (ordinary naïve participants, philosophical blog readers, and experts of Dennett's work). Our results showed that only the discrimination abilities of blog readers and experts were significantly above the chance rate of 20%[2] (blog readers 48%, experts 51%), even though lower than we hypothesized. In comparison, ordinary participants were near the chance rate of 20%. Given the expected improvement of future models, this suggests that probably even expertise in a domain will soon no longer provide a reliably way to distinguish machine-generated text from human-made text. Already now, experts on Dennett could, on average, only identify Dennett's answer half the time when presented with his answer alongside four answers from our fine-tuned language model.[3] These empirical results clearly show that human-machine discrimination abilities with respect to the linguistic output of LLMs are no longer a reliable criterion for identifying the linguistic outputs of LLMs beyond doubt.

---

[2] Chance rate is at 20% because we used a five-alternative forced choice task.

*2.2. Discrimination with the help of detection software*

One might think that if humans are unable to recognize the machine-generated text as machine-generated text, it should at least be possible to tell the difference beyond doubt using detection software. But at least with respect to the current state of research, even detection software cannot distinguish with 100% certainty between machine-generated and human-made text. Here, we seem to be at the beginning of an arms race between fraudsters and fraud detection.

Eric Mitchell and colleagues (2023) proposed a method called *DetectGPT* for deciding if a text passage was generated by a particular source model, for example, GPT-3. This method is based on the idea that if a text was generated by GPT-3, then this text has a high probability according to GPT-3, while human-written text does not have such a high probability from the point of view of GPT-3 (a nice explanation of this method can be found in the blog of Melanie Mitchell (2023)). Tests with several large language models showed that their method was able to distinguish between human-written and LLM-generated text in over 95% of the cases. However, 95% is not 100%, and also it is critical to note that the number of possible specific LLMs is constantly increasing, and since this method is specific to a particular model, the number of models that need to be tested may present a problem. One could argue that pretty much any LLM uses a similar neural network architecture and is trained with comparable training data. But, after all, one cannot rule out the possibility that there will be other LLMs in the future. Moreover, LLM users can manually set a preferred probability, to what extent this poses difficulties for this method would need to be tested.

It is important to keep in mind that current detectors for LLM-generated text commit two types of errors: false-negative (machine-generated text falsely judged to be written by humans)[4] and false-positive errors (human-generated text falsely judged to be machine-generated). False positives can be very harmful to humans, as I will describe in the next section. As long as we cannot exclude that such detectors falsely accuse humans of cheating, they should be used with caution and with the knowledge that their judgment could be false. This no longer indubitable distinctness leads to various other difficulties, which I will address in more detail below.

## 3. Ethical consequences

Due to the ever-increasing indistinguishability between machine-generated and human-generated texts, various ethical problems arise, especially in the age of electronic transfer. This begins with the no longer undoubtedly verifiable human authorship and continues with various types of fraud, such as a new form of plagiarism, but concerns also the violation of privacy rights, the possibility of circulating counterfeits of humans, and, last but not least, it enables the massive spread of misinformation.

Such consequences are certainly also supported by the hype that has taken place on social media. Investigating the first reactions to the release of LLMs like GPT-3 or ChatGPT, the voices expressing their astonishment and deep impression seem to be in the majority. However, right from the start, several scholars demonstrated how easy it is to expose LLMs (Marcus & Davis 2020). It seems that the critical voices have become more prominent only recently (Guardian Editorial 2023), even though the enthusiastic voices have not dried out. But even independent of a possible overestimation of the factual capabilities of LLMs, in the future, we will have to deal with ethical issues that arise primarily from the fact that we can no longer clearly distinguish machine-generated outputs from human-written outputs.

---

[3] In other domains, such as deep fake detection for audio (Groh et al. 2021; Müller et al. 2022) or differentiating human-made artwork from AI-generated artwork (Gangadharbatla 2022) also rather weak human discrimination capabilities were found.
[4] The first version of GPTzero (https://gptzero.me), for example, did evaluate the 40 machine-generated answers used in the experiment of Eric Schwitzgebel and colleagues as human-like.

*3.1. How to verify authorship*
Unless authors are directly monitored in the process of writing their texts, and it is ruled out that they can use LLMs in the writing process, it will no longer be possible to infer a human author beyond doubt on a textual basis. It could be that the accusation of passing off a machine-generated text as one's own can no longer be dispelled in the last instance. This means that it could become difficult to prove oneself beyond doubt as the author of a text when submitting a paper.

Assuming that also in the future, neither humans nor detection software can reliably distinguish between machine-generated and human-written text, we have to be prepared to deal with false positives and false negatives when trying to recognize a human-written text. From that point of view, it is conceivable that an author submitting a paper could be falsely accused of having delivered a machine-generated text, and conversely, it is equally possible for a machine-generated text could pass as human-generated. As LLMs continue to advance, it may even become common to submit newspaper articles or even articles to scientific journals in which much of the content is produced by machines. For sure, it is expected that students could make use of LLMs to let them produce text for their essays, and their teachers will not be able to recognize whether the students delivered self-made texts (Herman 2022; Hutson 2022; Huang 2023; Marche 2022; Peritz 2022; Sparrow 2022). Universities might turn back to in-person exams to reassure the authorship of their students. However, this is not possible in all cases where authorship matters, especially with respect to the mass of electronically distributed texts. How new chains of trust can be established will be a challenge for future societies.

*3.2. New forms of plagiarism*
Since language models create novel sentences – they are not parrots – it is unlikely that their output will lead to any results when using standard plagiarism checkers. Using plagiarism checkers, similarity thresholds below 10%-15% are considered ordinary for non-plagiarized work (Mahian et al. 2017). Schwitzgebel and colleagues (2023) investigated whether fine-tuning GPT-3 on Dennett's works might have led the model to be overtrained[5] so that it simply parroted sentences or multi-word strings of texts from Dennett's corpus. To verify that the fine-tuned model was indeed producing new texts, we used the Turnitin plagiarism checker to check for "plagiarism" between the machine-generated outputs and the Turnitin corpus supplemented with the works that were used as the training data. Turnitin reported an overall similarity of 5 % between the GPT-3 generated answers and the comparison corpora, and none of the passages were flagged as similar to the training corpus. Also, the search for matching text strings between the GPT-3 responses and the training corpus revealed no matches, except for some stock phrases favored by analytic philosophers. Thus, the machine-generated text does not simply plagiarize its training data word for word but generates novel – albeit stylistically and philosophically similar – content.

Nevertheless, it is at least arguable whether outputs of fine-tuned language models should be considered plagiarism because they sort of 'borrow' ideas from their training data with which they were fine-tuned without acknowledging the original author.

*3.3. Violation of copyright rights and privacy*
Another issue concerns the intellectual property of the persons who wrote the text with which LLMs are trained. For example, OpenAI's GPT-3 was trained on hundreds of billions of words of text (499 billion tokens)[6] from Common Crawl, WebText, books, and Wikipedia (Brown et al. 2020). And especially fine-tuned models like the one we fine-tuned on the corpus of Daniel Dennett have to put up with the question of whether this is a fair use of someone else's intellectual property to use their works in creating an LLM without asking permission. For this reason, we asked Daniel Dennett for the explicit permission before fine-tuning our LLM on his texts, and we agreed that he has the final say when it comes to the question of who may use this language model and which outputs are published.

---

[5] Being overtrained is an issue regarding neural networks. However, running four epochs of fine-tuning is a standard recommendation from OpenAI, and in most applications, four epochs of training do not result in overtraining (Brownlee 2019).

[6] A token is a sequence of commonly co-occurring characters, with approximately ¾ of an English word per token on average.

It goes without saying that we always explicitly label the outputs of our model as machine-generated output.

However, in the case of already deceased individuals, it is not possible to ask for permission (for a review regarding the potential use of personal data of deceased persons, see Nakagawa & Orita 2022).

To date, copyright laws regarding the use of copyrighted text as training data for fine-tuning language models have not yet been clarified (see Government UK consultations 2021).

And the question of whether output from LLMs could be protected by copyright has not yet been conclusively resolved. According to a report in The Verge (Robertson 2022), the U.S. Copyright Office recently rejected a request to grant copyright to a work of art to an AI because, in their view, it was a necessary standard for protection that the work of art contain at least elements of "human authorship."

*3.4. Counterfeits of people*

Furthermore, it is a matter of fact that people leave behind a lot of private data that are not secured at all. LLMs can be fine-tuned on all kinds of additional training data. This could concern any information a person has shared on social media. For example, one could use all kinds of data available about someone's life, e.g., in social media, blogs, websites, online stores, and search engines, to create a digital replica that could convincingly imitate some of that person's behavior (Karpus & Strasser under review). A striking illustration of such a replica, albeit a fictional one, can be found in the episode "Be right back" of the television series Black Mirror (Brooker 2013). This episode inspired Eugenia Kuyda to feed all the saved online conversations she had with a deceased friend into an AI-powered system to create a chatbot version of her friend. Subsequently, a public application called *The Replika* was created (see https://replika.com; Murphy 2019).

The extent to which existing LLMs, tools, and chatbots based on more or less private data violate copyright rights and rights concerning privacy seems to me to be an open question that we should urgently address. Even if the science fiction story depicted in the Black Mirror episode is still too futuristic in many aspects in the context of today's technology, it is worth considering how revealing the data we leave behind on social media, search engines, online stores, and other platforms already is about our character traits concerning our likes, dislikes, desires and other features of our personality. Besides the scandals involving Cambridge Analytica in the UK which demonstrated the richness of the data available about our lives on social media platforms, there is an insightful art project by the Berlin-based artist collective Laokoon (https://www.madetomeasure.online/en). Their investigative project 'Made to measure' explored the question of how far one can get in constructing a doppelganger of a person using data available online about that person's life. Using anonymized data of a person's Google search history from five years of that person's life, they retraced the life of that person and re-enacted that life in a film. When confronting the data donor, who was contacted after processing the data with the help of a personalized Instagram message, it turned out that the reconstruction of this person's life was amazingly accurate in many aspects (for a more detailed description, see Karpus & Strasser under review).

If digital replicas start to speak on behalf of the person out of which data they were constructed, they could be considered as a counterfeit of this person. In other words, if outputs of LLMs were presented as a quotation or paraphrase of positions of existing persons, this would constitute counterfeiting (Dennett as interviewed in Cukier 2022).

With the help of increasingly sophisticated language models, we will be able to create fakes of people that are difficult to distinguish from their originals. Just as we are able to create counterfeit money, it is conceivable that in virtual communication, for example, one can create the appearance of interacting with a real person that turns out to be a fake. Such deep fakes do not have to be restricted to linguistic output; they can also be enriched with visual and auditory imitations. In Germany, for example, a mayor was made to believe that she was interacting with Vitali Klitschko in a zoom-call (Hoppenstedt 2022). Another example, again from the field of art, that illustrates the potential of counterfeiting is the art project "Chomsky vs. Chomsky" (Rodriguez 2022). It is important to note that while it was made clear from the outset that this is an artifact and not the real Chomsky, this project nevertheless shows how disturbing a digital replica can be. This art project presents a virtual version of Noam Chomsky – a location-based, Mixed Reality (MR) experience that draws not only on Chomsky's texts but also on recorded lectures. Thereby, this project offers the experience of asking questions orally in virtual reality and receiving an audio response whose sound is almost indistinguishable from the recordings of the real Chomsky.

No doubt, it should be against the law to present a conversational AI without making it clear that it is an AI and not a human. If an LLM counterfeits people, the creator and the users of this LLM are guilty of a crime. However, until now, we do not have the legal basis to prosecute such crimes.

*3.5. Spread of misinformation, nonsense and toxic language*

Stepping back from the emotional and overwhelming judgments shared in social media, it is an important and serious question to investigate to what extent we can trust machine-generated assertions.

The danger of mistakenly trusting GPT-3 is particularly evident in the health sector. We should be clear about the fact that within this domain, we are nowhere near any application where GPT-3 could provide reliable help in any sense. It lacks the scientific and medical expertise that would make it useful for any medical Q&A, as it can be very wrong, and this is not viable in healthcare. For example, in a test where GPT-3 responded to mental health problems, the AI advised a simulated patient to commit suicide (Daws 2020). Nevertheless, despite the warning from OpenAI it is probably to be expected that such applications will be developed.

It is a matter of fact that LLMs based on a transformer architecture with a statistical self-attention mechanism – machines that calculate the probability of words appearing in the context of other words – have severe limitations regarding reliability. This becomes evident, for example, when LLMs make self-contradictory statements. It is quite possible that if an LLM receives the same question as a prompt several times, it will respond with very different answers which are contradicting each other. In contrast, a standard calculator will always give the same answer, for example, it will always 'claim' that 2+2 equals 4, whereas LLMs are able to 'claim' that 2+2 equals 4 in one instance and that 2+2 equals 5 in another. For this reason, it is important that a balanced assessment of the performance of LLMs should not rely solely on cherry-picked, mind-boggling outputs.

Moreover, the better the models become, the more difficult it becomes to distinguish machine-generated linguistic outputs from human-made utterances, and at the same time, the risk of misuse increases. For example, LLMs can play a weighty role in spreading misinformation (Marcus 2022). Since LLMs are inherently unreliable, they do make severe mistakes in reasoning and facts. This unreliability is due in part to the fact that LLMs build models of word sequences based on how humans use language rather than models describing how the world works. Although it can be concluded from this that many machine-generated linguistic results are correct because human language often reflects facts in the world. But at the same time, it follows that the accuracy of LLM's statements is, to some extent, a matter of chance because, unlike humans, machines do not use language to refer to the world. An example that can illustrate the production of misinformation and nonsense is the brief presence of Galactica. This LLM was created to write plausible-sounding academic papers. However, it was taken offline a few days after its release due to harsh criticism amounting to the claim that Galactica produces vaguely-plausible-sounding-but-ultimately-nonsensical academic papers (Al-Sibai 2022; Taylor et al. 2022). Likewise, I suppose that it is to be expected that applications that aim to assist search applications like ChatGPT with respect to Bing will soon be taken offline (Rogers 2023).

Lately, the most discussed LLM has been ChatGPT. Unfortunately, ChatGPT has a tendency to hallucinate – it produces statements that sound plausible but are simply false. It is able to invent references of papers that were never been written, it can make up historical dates, and it commits severe failures regarding the solution of logical problems. Furthermore, LLMs lack what we would call 'common sense' in the human case. It is a serious problem that LLMs can produce sentences that are simply not true (for a repository of errors made by LLMs, see Davis et al. 2023, Marcus & Davis 2023). To date, LLMs have no reliable mechanisms for verifying the truth of their statements, and this is effectively a springboard for the mass production of misinformation. For example, Gary Marcus (2023) reports that the independent researcher Shawn Oakley has shown how easy it is to get ChatGPT to produce misinformation that it even backs up with fictional studies. This is especially troubling as ChatGPT adopts an authoritative tone.

Another problem with LLMs is that their outputs depend on their training data. This means that if they are not constantly retrained, they quickly become obsolete in terms of up-to-date information. ChatGPT lacks 'knowledge' of events that occurred after 2021. They can make prophecy-like statements about events that may have happened in the meantime, but these statements lack any relation to our reality.

Because of their limited reliability, LLMs require human supervision, but this leads to another critical ethical issue. OpenAI strives to filter out toxic content (e.g., sexual abuse, violence, racism, sexism, etc.), which is a good goal in principle, but the implementation is not perfect and is also highly questionable. To flag toxic data produced by LLMs, one needs humans, and this work is traumatic in nature. According to a *TIME* investigation, OpenAI used outsourced Kenyan laborers earning less than $2 per hour to make ChatGPT less toxic (Perrigo 2023). Apart from the fact that this practice is ethically questionable, it is also not ultimately successful from a technical point of view.

### 4. How to handle the epistemological crisis

Due to all potential deep fakes, there is an epistemological crisis to be expected, and people will need to look out for what they take as representing a real person. Avoiding that we get too suspicious and paranoid, we need laws for how AIs present themselves, and we will probably have to develop new strategies for identifying our counterparts as humans.

One helpful measure would be to create a legal basis for requiring machine-generated output to be labeled as machine-generated text as a matter of principle. This is addressed, for example, by a recent proposal, the so-called AI-act of the European Commission (2021), which requires labeling for anything that might be mistaken for human interaction. Such regulations could help mitigate the risk that LLMs will be used to contribute to a huge spread of misinformation.

One way to label machine-generated text could be accomplished through the use of digital watermarks (Wigger 2022). Kirchenbauer and colleagues (2023) have suggested that one way to do this would be to require the creators of LLMs to add a watermark signal to each generated text passage that cannot be easily removed by simply modifying the text, and to provide open-source software for watermark detection. This sounds good at first glance, but one cannot assume that all LLM creators will adhere to it, and of course, it is also possible to fool watermark detectors. Again, it is likely that an arms race will develop here between fraudsters and those who want to mark LLM's outputs recognizably.

And as described above, so far, there is no completely reliable method for detecting AI-generated text. So, bans cannot be enforced proactively, which means that one has to rely on human help. For example, a conference has banned the use of machine-generated text in submissions. However, in the end, only submissions that are deemed suspicious to other scientists were examined (Vincent 2022). It seems as if we are not prepared for the emergence of such disruptive and novel technologies.

### 5. LLMs as thinking tools

All of this is not to say that positive applications of LLMs are not conceivable. Many applications that we use on a daily basis are based on models that have been trained to predict the next word or words in a text based on the preceding words, e.g., it is part of the technology that predicts the next word you want to type on your mobile phone allowing you to complete the message faster. Likewise, the increased quality of translation software is indeed a helpful tool, although it is still advisable not to release the translation without human review.

In other domains, such as game-playing AIs, neural network architectures have led to success (Campbell 2002; Brown & Sandholm 2019; Silver et al. 2016, 2018). And impressive results can also be pointed to in scientific fields, such as a program for lung cancer screening (Ardila et al. 2019), or AlphaFold, that is predicting protein structure (Jumper et al. 2021), and AlphaTensor, that is discovering novel matrix multiplication algorithms (Fawzi et al. 2022). However, it is important to keep in mind that the performance of these neural networks is tied to clearly definable goals. The output generated by machines can be evaluated and verified on the basis of clear criteria. A victory in a chess game is clearly defined, and the individual steps are also subject to a set of rules. However, when we move into the realm of communication with human language, it is not always clear whether an answer to a question and its justifications meet our requirements for comprehension, rational thinking, and the like.

Considering the potential of large language models and assuming that technology will continue to advance, it is conceivable that language models will soon produce results interesting enough to serve as a valuable resource for human researchers. In other domains, computer programs are able to generate music in the style of a particular composer (Hadjeres et al. 2017; Daly 2021; Elgammal 2021) or create all kinds of images (DALL-E). Even if not all outputs are reliable or interesting, selected outputs seem to have significant musical or artistic value. A composer or artist could produce many outputs, select the most promising, edit them slightly, and present them as original works. As a matter of fact, there was already a case where an AI-produced picture did win a competition (Metz 2022).

In this way, language models could become thinking tools that people use. However, it is important that the users are experts in their domains and remain able to validate the outputs. For example, a researcher could fine-tune a language model with certain corpora and then generate outputs they can use as inspiration for further ideas. However, when using language models as thinking tools, one must be careful not to rely too heavily on them as deep learning networks always have reliability issues (Alshemali & Kalita 2020; Bosio et al. 2019). A user with insufficient expertise might mistakenly assume that all results from a large language model fine-tuned to an author's work reflect the author's actual views (Bender et al. 2021; Wedinger et al. 2021). The use of such language models will not be able to replace reading the original works (see Steven & Iziev 2022), but they may eventually become a helpful tool for humans creating text.

According to a whitepaper published by Lionbridge (2023), ChatGPT can help with translation, terminology, style guides, content classification, post-editing, content analysis, and creating working code. However, it can only help – one can never rely on an LLM to say true things or know what is right or wrong, outputs of all LLMs are unreliable. Therefore, it remains up to humans to decide what makes sense and what is true or false. AI can be used to improve, polish, edit, or write texts, but it will still be up to humans to judge their value.